\documentclass{PoS}

\usepackage{graphicx}

\long\def\symbolfootnote[#1]#2{\begingroup%
\def\thefootnote{\fnsymbol{footnote}}\footnote[#1]{#2}\endgroup} 
\def\apj{ApJ}
\def\aap{A\&A}
\def\mnras{MNRAS}
\def\nat{Nature}

\newcommand{\nustar}{\textit{NuSTAR}}

\newcommand{\xmm}{{\it XMM-Newton}}

\newcommand{\Lya}{Ly$\alpha$}

\title{Relativistic spectroscopy of the extreme NLS1 IRAS 13224-3809}

\ShortTitle{Relativistic spectroscopy of IRAS~13224}

\author{\speaker{M. L. Parker}\\
        European Space Agency (ESA), European Space Astronomy Centre (ESAC), E-28691 Villanueva de la Ca\~{n}ada, Madrid, Spain\\
        E-mail: \email{mparker@sciops.esa.int}}

\author{W. N. Alston, D. J. K Buisson, A. C. Fabian, J. Jiang, C. Pinto\\
		Institute of Astronomy, Madingley Road, Cambridge, CB3 0HA, UK }
\author{E. Kara\\
		Department of Astronomy, University of Maryland, College Park, MD 20742-2421, USA}
\author{A. Lohfink\\
		Department of Physics, Montana State University, Bozeman, MT 59717-3840, USA}

\abstract{The narrow line Seyfert 1 (NLS1) IRAS~13224-3809 is the most X-ray variable active galactic nucleus (AGN), exhibiting 0.3--10~keV flux changes of over an order of magnitude within an hour. We report on the results of the 1.5~Ms 2016  \xmm/\nustar\ observing campaign, which revealed the presence of a 0.24c ultra-fast outflow in addition to the well-known strong relativistic reflection. We also summarise other key results of the campaign, such as the first detection of a non-linear RMS-flux relation in an accreting source, correlations between outflow absorption strength/velocity and source flux, and a disconnect between the X-ray and UV emission. Our results are consistent with a scenario where a disk wind is launched close to the black hole, imprinting absorption features into the spectrum and variability.}

\FullConference{Revisiting narrow-line Seyfert 1 galaxies and their place in the Universe - NLS1-2018\\
		9-13 April 2018 \\
		Padova Botanical Garden, Italy}

\begin{document}

\section{Introduction}

IRAS~13224-3809 (hereafter IRAS~13224) is the most variable AGN in the CAIXA catalogue \cite{Ponti13}, closely followed by the very similar source 1H~0707-495. It shows very strong broad emission lines, the signature of X-ray reflection from the inner accretion disk \cite{Fabian13_iras, Chiang15}, and reverberation lags produced by the time delay between the X-ray source and the disk \cite{Kara13_iras}. In 2016, IRAS~13224 was continuously observed for 1.5~Ms with \xmm, and 500~ks with \nustar\ (PI Fabian). This has resulted in a truly exceptional dataset for the study of accretion physics and relativistic phenomena around supermassive black holes, and has directly lead to several major discoveries.

\section{Time-averaged spectroscopy}

\begin{figure}
    \centering
    \includegraphics[width=0.8\linewidth]{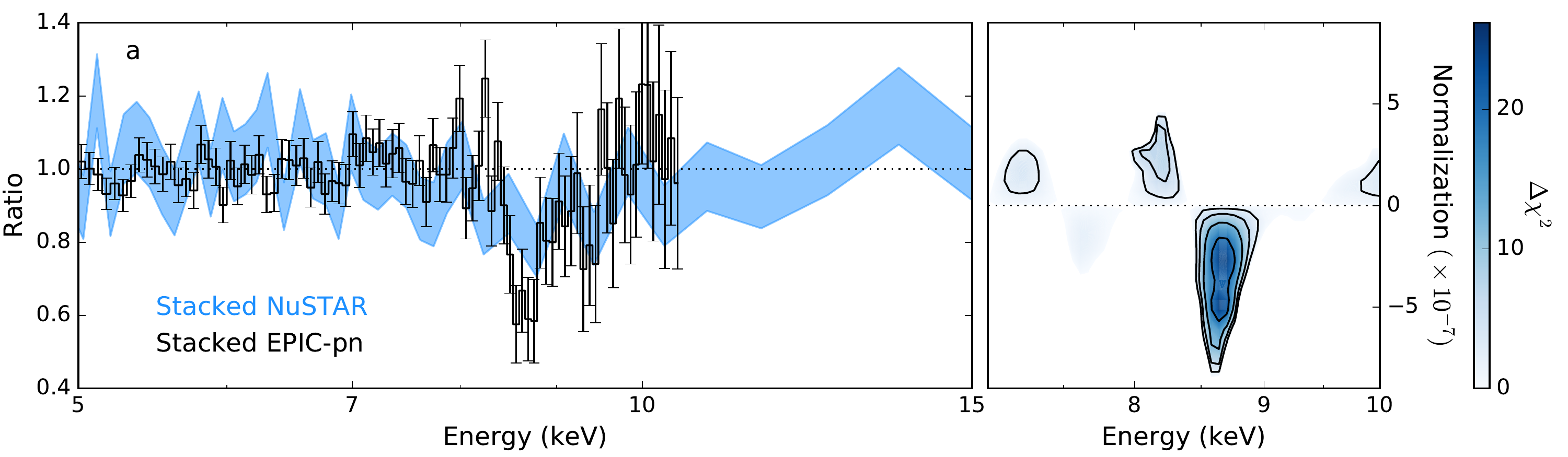}
    \caption{Reproduced from \cite{Parker17_nature}. Left: Time-averaged \xmm\ and \nustar\ residuals, showing the presence of a strong absorption line at $\sim8.6$~keV. Above 7~keV, the only absorption lines strong enough to give a unique absorption feature are the Fe\textsc{xxv} and Fe\textsc{xxvi} lines.}
    \label{fig:nature_resids}
\end{figure}

Investigation of the time-averaged \xmm\ EPIC-pn and \nustar\ FPM spectra showed a strong absorption feature at $\sim8.6$~keV, due to a blend of Fe\textsc{xxv/xxvi} \cite{Parker17_nature} at $v=0.24$c (see Fig.~\ref{fig:nature_resids}). Formally detected at $>5\sigma$, this feature is likely produced by an ultra-fast outflow (UFO), launched from the accretion disk at high-Eddington accretion rates. Alternatively, these velocities are also produced on the surface of the disk, and an absorbing layer on the disk could therefore potentially give the same absorption features.

\begin{figure}
    \centering
    \includegraphics[height=0.4\linewidth]{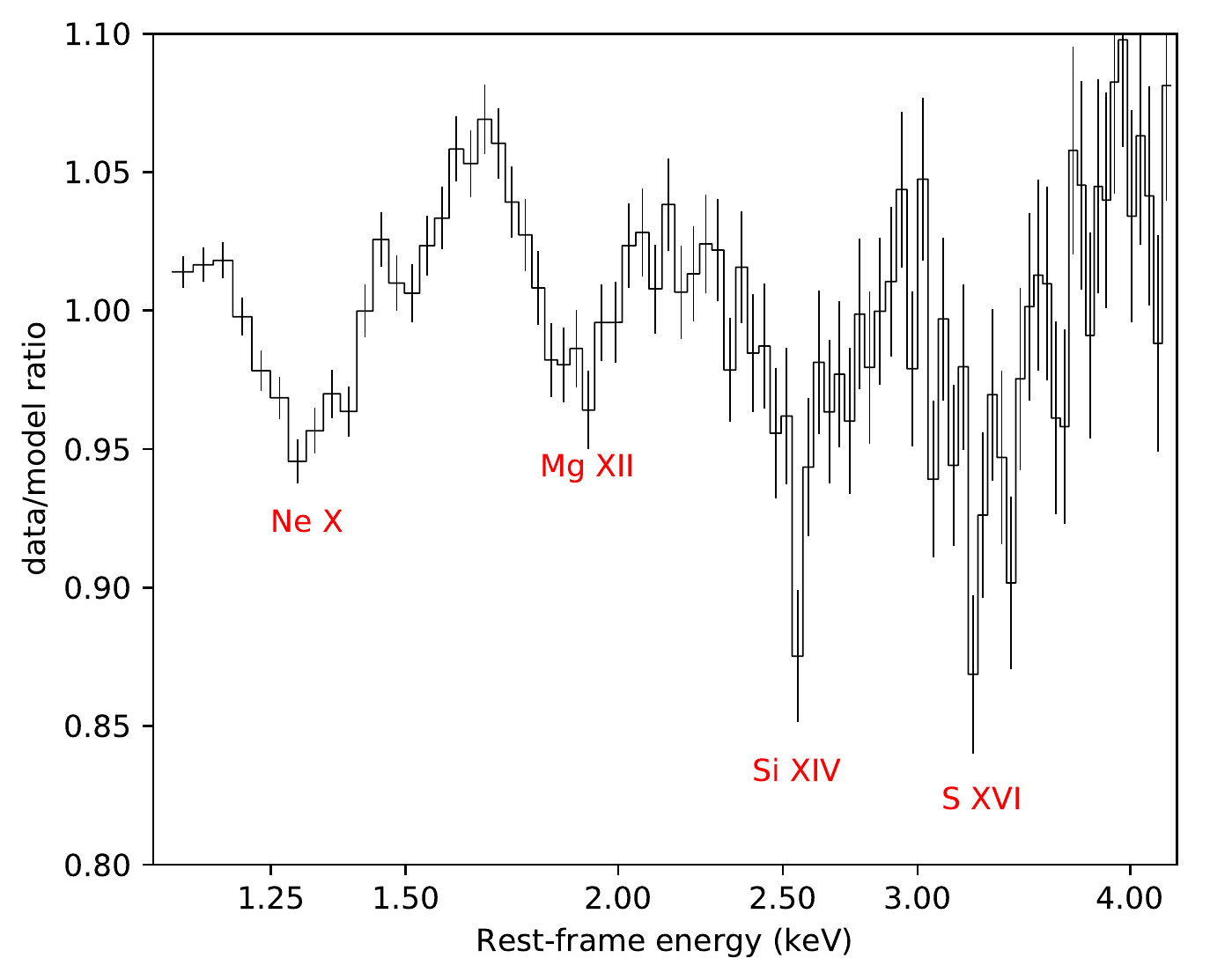}
    \includegraphics[height=0.4\linewidth]{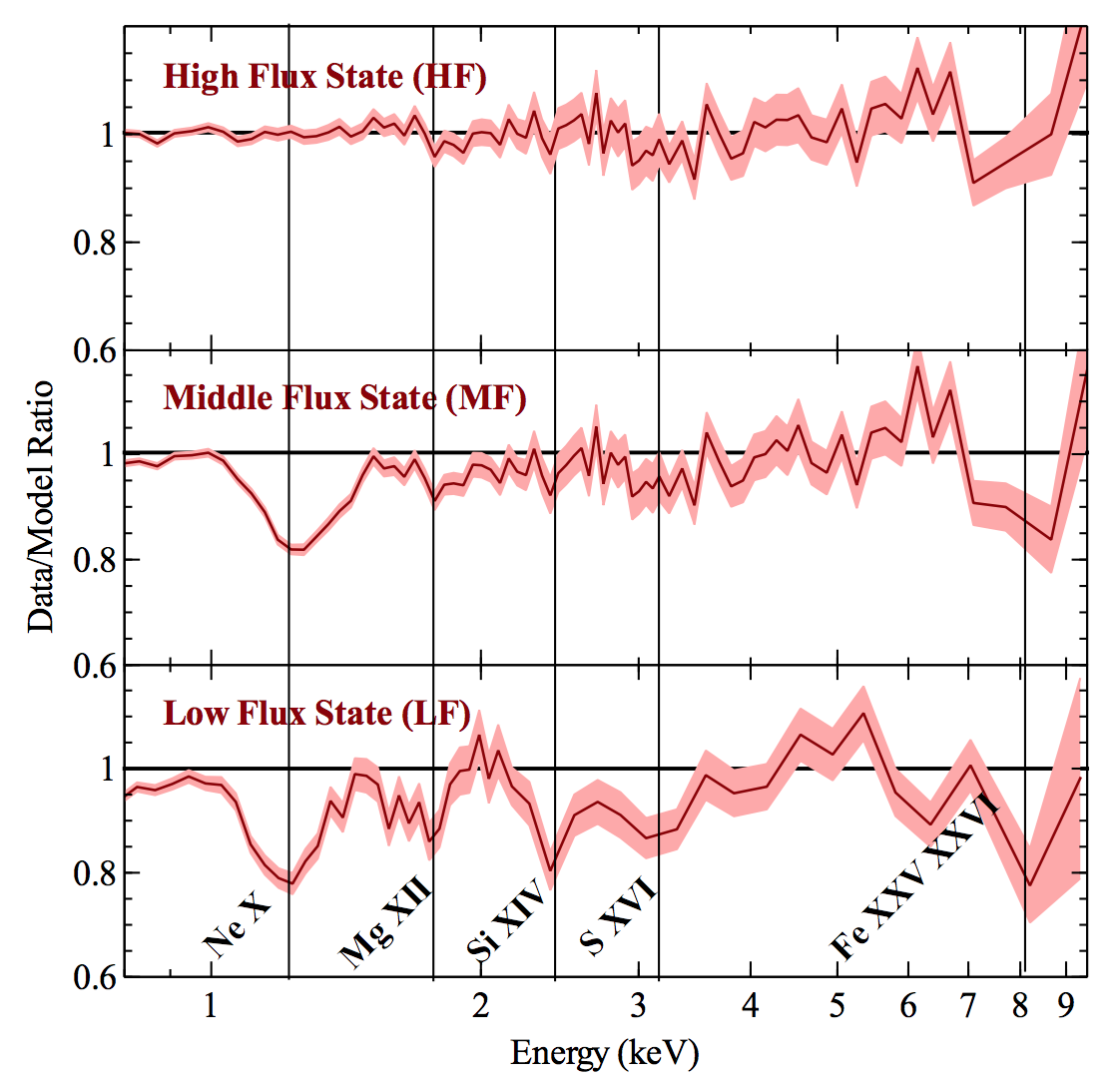}
    \caption{Left: intermediate energy absorption lines from the EPIC-pn time-averaged spectrum. Right: Flux dependence of the UFO, reproduced from \cite{Jiang18}.}
    \label{fig:intermediatelines}
\end{figure}

Fitting the Fe~K band with a combination of relativistic reflection and ionized absorption gives fit parameters that are consistent with those found by previous authors \cite{Fabian13_iras, Chiang15}, indicating that those results were not strongly affected by not including the UFO in their fits.
Following on from this, a detailed inspection of the EPIC and RGS spectra at lower energies revealed additional absorption features from O\textsc{viii} Ly$\alpha$ and Ly$\beta$, and the Ly$\alpha$ lines of Ne\textsc{x}, Mg\textsc{xii}, Si\textsc{xiv} and S\textsc{xvi} (see the left panels of Fig.~\ref{fig:intermediatelines} and Fig.~\ref{fig:pca}).

The presence of both relativistic reflection and an ultra-fast outflow in an AGN is interesting, as it more strongly constrains the UFO geometry \cite{Parker18_geometry}. Additionally, given that the accretion rate is near-Eddington (see discussion of this in \cite{Jiang18}), it is possible that the UFO removes enough energy from the accretion flow before it reaches small radii that the disk stays cool enough to produce strong relativistic reflection. In this scenario, it is therefore the UFO that ensures the presence of a disk capable of producing strong relativistic reflection signatures while at high Eddington rates. Alternatively, it is possible that the absorption features could be produced in an absorbing layer on the disk, where these velocities occur naturally.

\section{Flux-resolved spectra}
\begin{figure}
    \centering
    \includegraphics[width=0.9\linewidth]{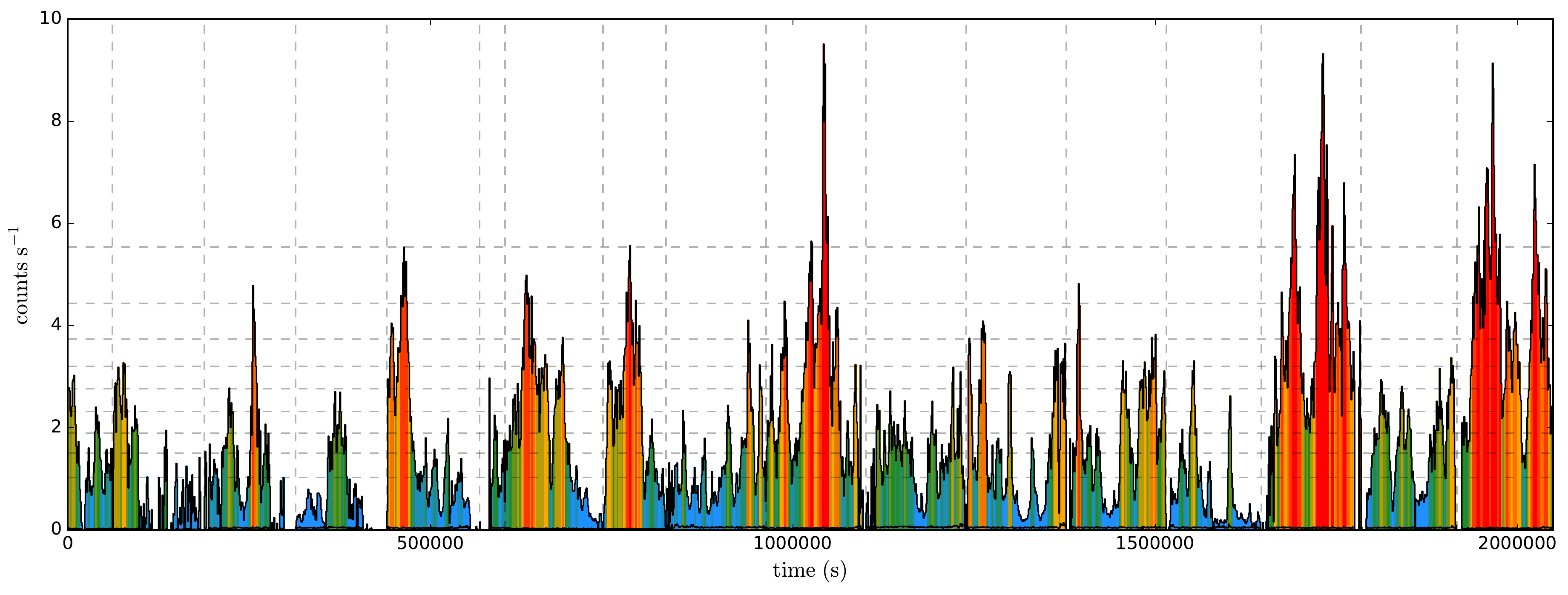}
    \caption{Lightcurve of the 2016 observing campaign. Vertical dashed lines indicate different observations, horizontal lines and colours indicate different flux levels.}
    \label{fig:lightcurve}
\end{figure}

\begin{figure}
    \centering
    \includegraphics[height=0.4\linewidth]{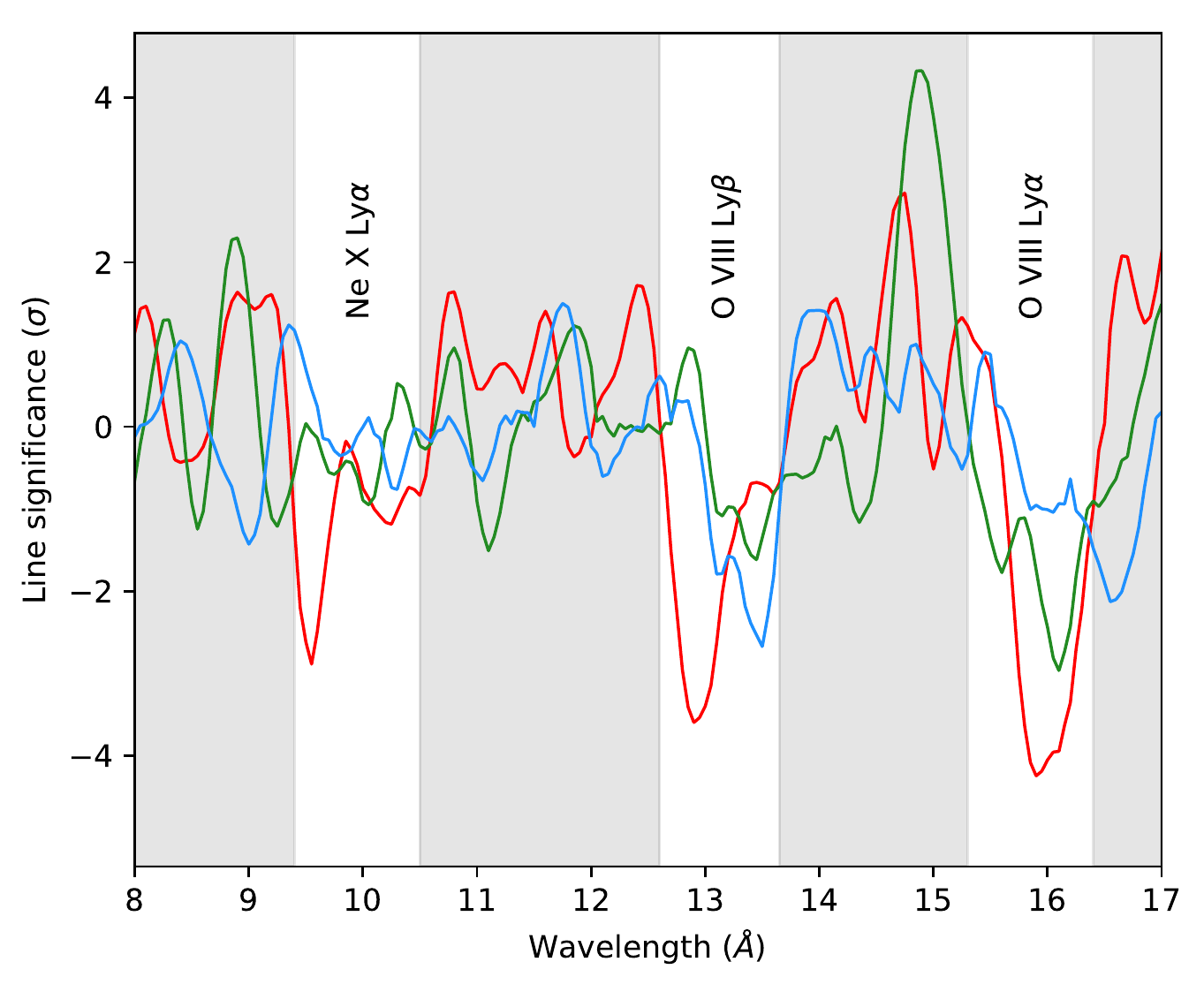}
    \includegraphics[height=0.4\linewidth]{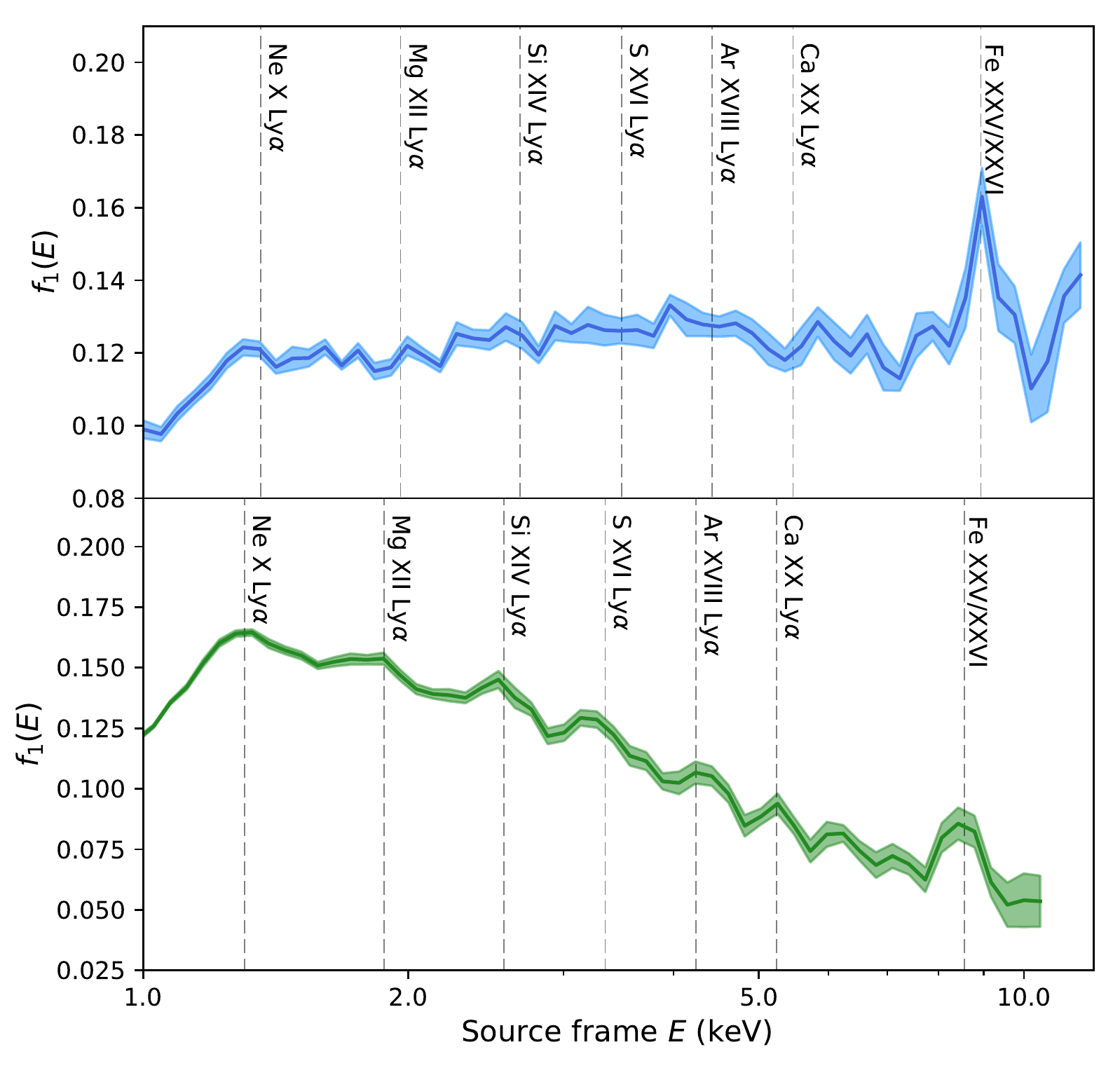}
    \caption{Left: Line significance in the RGS, for low, medium and high fluxes (red, green and blue, respectively). Three flux-dependent features are visible, corresponding to the predicted energies of the Ne and O \Lya\ lines and the O Ly$\beta$ line. Right: PCA spectra of IRAS~13224 (bottom) and PDS~456 (top) showing enhanced variability in UFO absorption lines. Note that PDS~456 does not have the lower energy lines, due to the higher ionization of the gas.}
    \label{fig:pca}
\end{figure}

\begin{figure}
    \centering
    \includegraphics[height=0.38\linewidth]{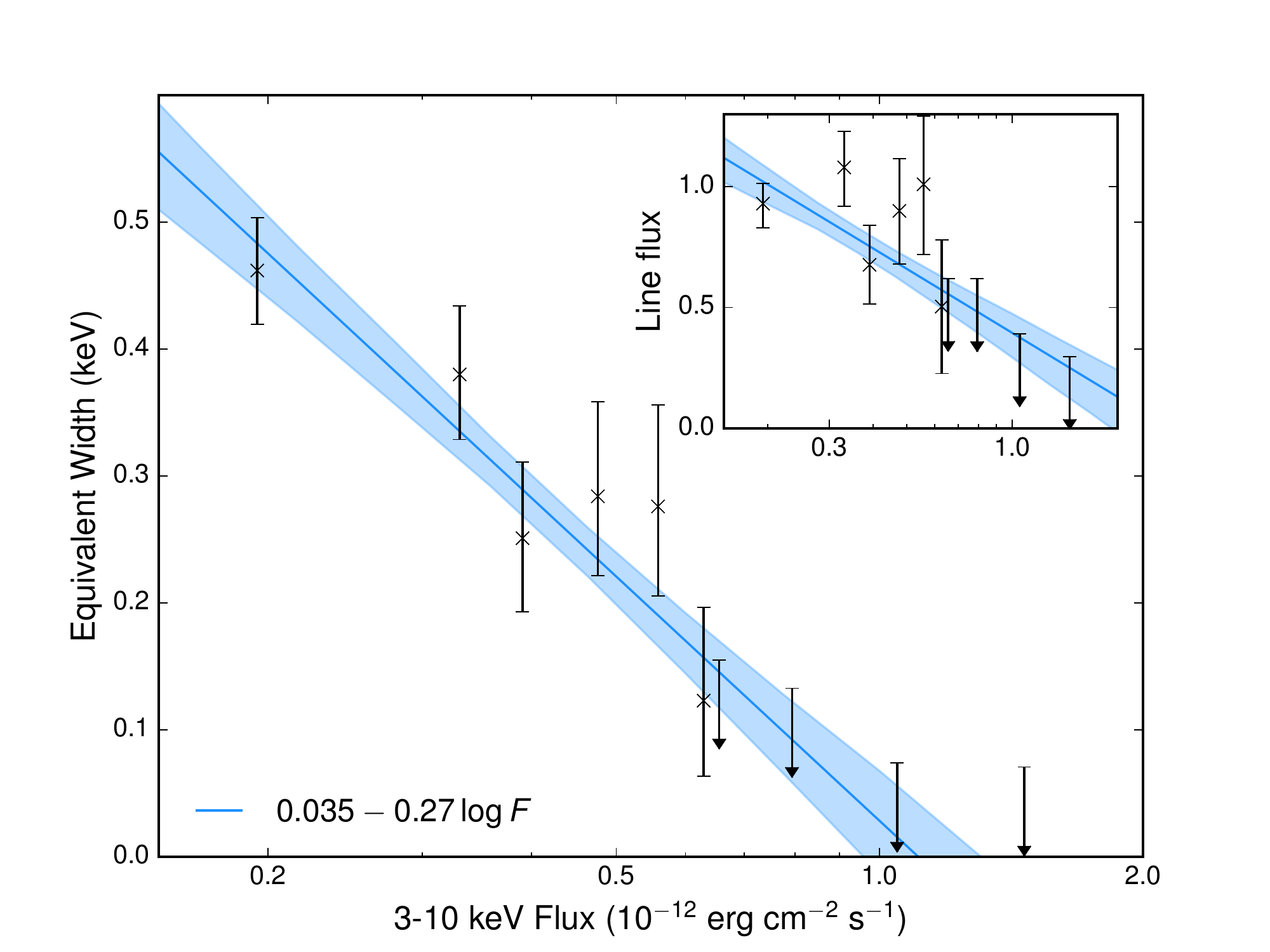}
    \includegraphics[height=0.35\linewidth]{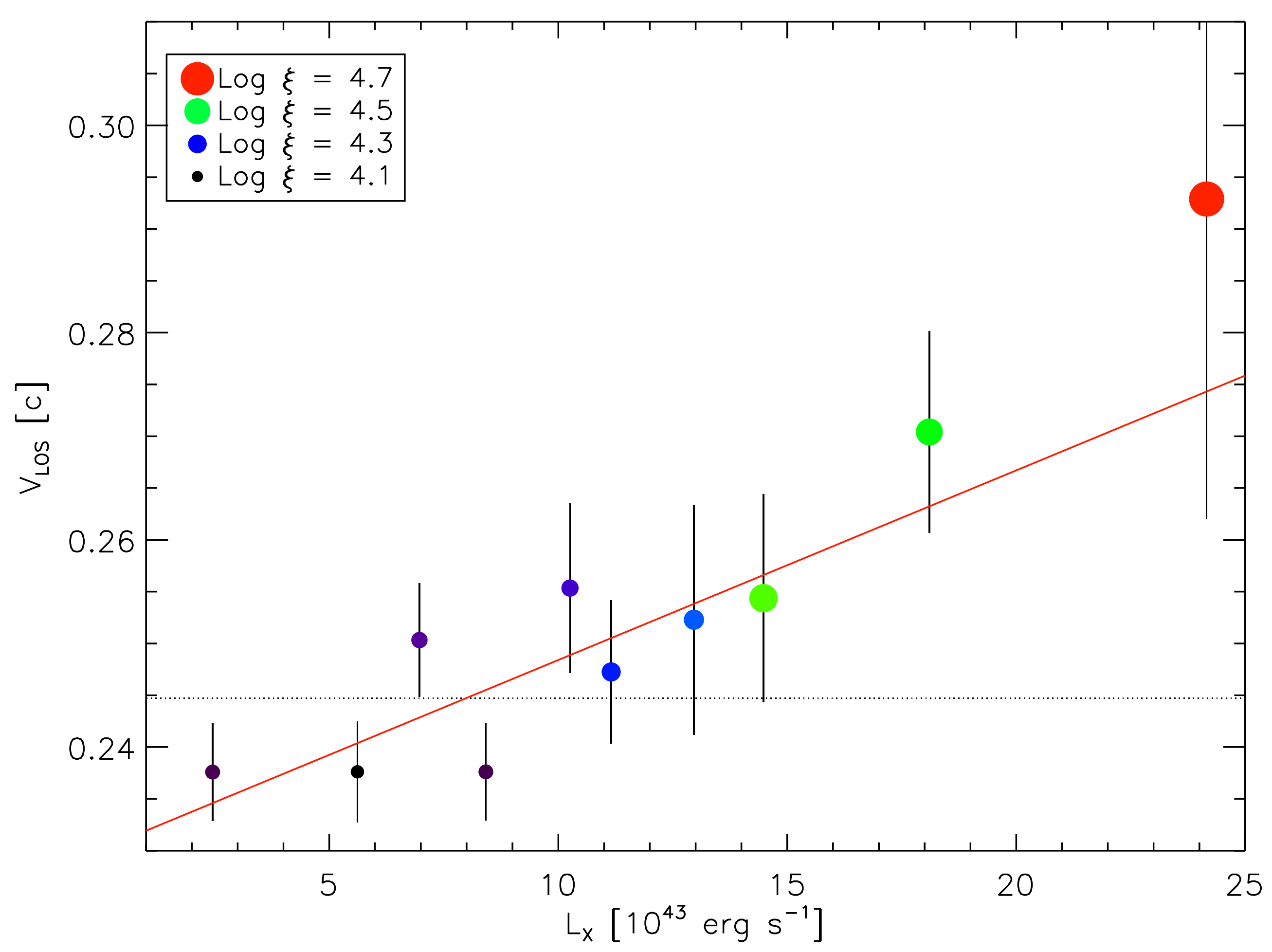}
    \caption{Left: correlation between 3--10~keV source flux and the equivalent width of the Fe\textsc{xxv/xxvi} absorption line, from \cite{Parker17_nature}. Right: Correlation between source luminosity and UFO velocity, from \cite{Pinto18}}
    \label{fig:correlations}
\end{figure}

Because of the extreme flux variability (see section~\ref{sec:variability}), it is easy to split the dataset into flux-resolved spectra. In Fig.~\ref{fig:lightcurve} we show the EPIC-pn lightcurve of the 2016 campaign, divided into different flux levels so that the total number of counts in each level is approximately the same. By calculating spectra for three different flux levels, it is immediately obvious that the absorption lines show a strong flux dependence \cite{Parker17_nature,Jiang18} (Fig.~\ref{fig:intermediatelines}, right, Fig.~\ref{fig:pca}, left). A closer investigation, splitting the data in 10 flux levels, gives a clear correlation between equivalent width and flux \cite{Parker17_nature}, and velocity and flux \cite{Pinto18} (Fig~\ref{fig:correlations}). The equivalent width trend is likely produced by photoionization of the outflowing gas, fully ionizing the material and removing the lines.

\section{Variability and timing}
\label{sec:variability}
The time-domain variability properties of such a variable AGN are also of great interest. Using simple variability spectra, we demonstrated that the UFO absorption lines produce variability `spikes', where the fractional variability is enhanced, due to the anticorrelation between equivalent width and flux \cite{Parker17_irasvariability,Parker18_pds456}. 
In Fig.~\ref{fig:pca}, right, we show the principal component analysis (PCA) spectra \cite{Parker15_pcasample} for IRAS~13224 and another AGN with a strong UFO, PDS~456. In both cases, the UFO lines are highlighted clearly in the variability spectrum, and the different ionization states of the two outflows are apparent. This is potentially a powerful new method for finding such outflows, as it does not require any time-consuming and model-dependent spectral fitting, can be used on the whole band simultaneously, and is less susceptible to background contamination than conventional methods.

\begin{figure}
    \centering
    \includegraphics[height=0.35\linewidth]{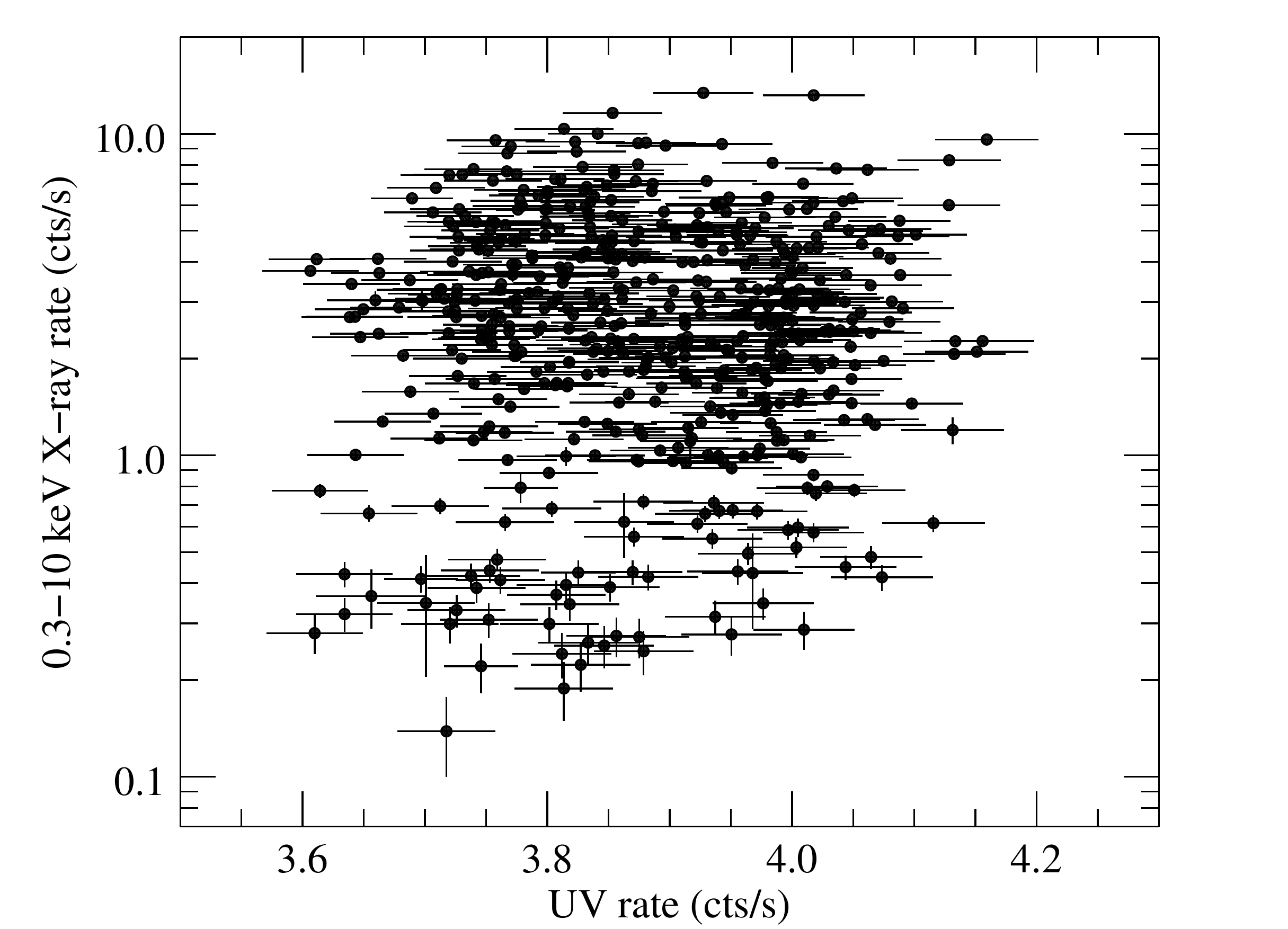}
    \includegraphics[height=0.38\linewidth]{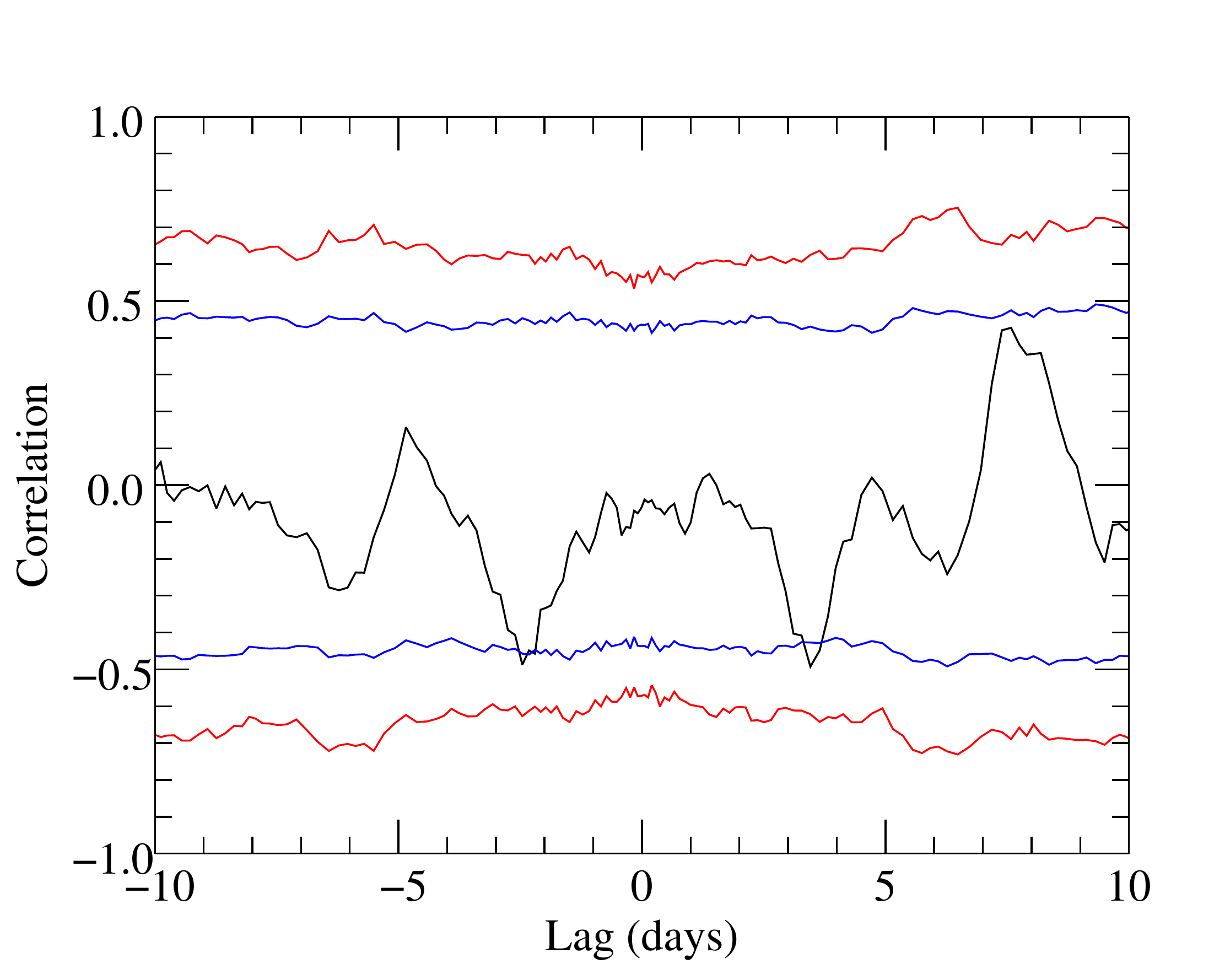}
    \caption{Reproduced from \cite{Buisson18}. Left: X-ray against UV flux. The points appear completely unrelated. Right: Discrete correlation function of X-ray and UV light curves, 
    with 90\% and 95\% confidence limits around zero correlation. There is still no significant correlation found, even with a lag allowed between emission in the two bands.}
    \label{fig:variability_buisson}
\end{figure}

In many AGN there is a delay between the X-ray emission and the UV emission. This is usually interpreted as being due to the X-rays heating the outer disk, causing an increase in the number of UV photons produced \cite{Shappee14, McHardy14, Buisson17}. However, in the case of IRAS~13224-3809 there is apparently no connection between the X-rays and the UV (Fig.~\ref{fig:variability_buisson}). Indeed, the UV is no more variable than the average AGN, despite the extreme X-ray variability. This suggests that the UV emitting region is not seeing the X-ray variability, potentially because the outer disk is blocked from the X-ray source by the base of the outflowing disk wind.

\begin{figure}
    \centering
    \includegraphics[width=0.4\linewidth]{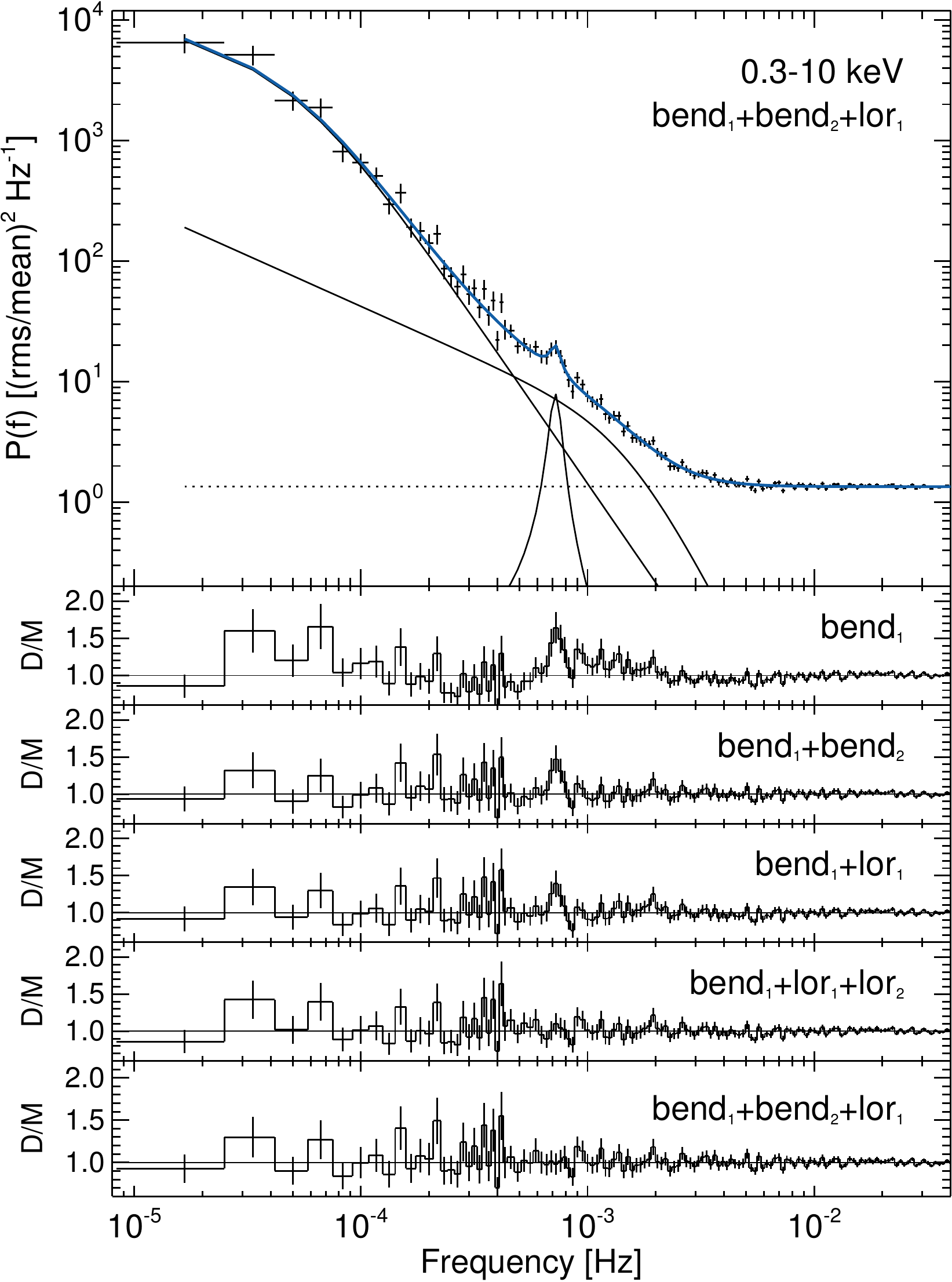}
    \includegraphics[width=0.4\linewidth]{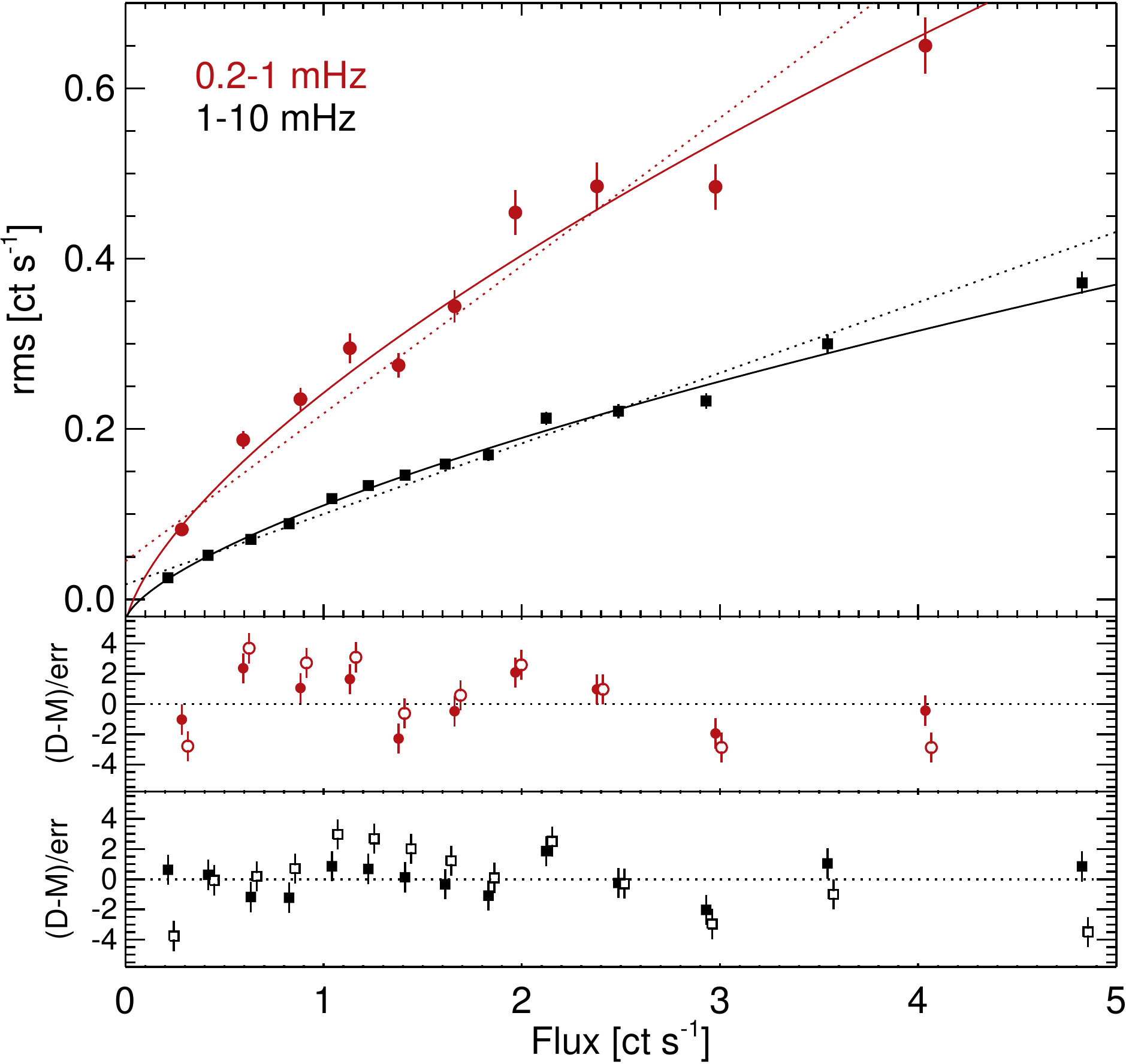}
    \caption{Reproduced from \cite{Alston18}. Left: X-ray power spectrum of IRAS~13224. Right: RMS-flux relation, showing a deviation from the well known linear relation.}
    \label{fig:timing_alston}
\end{figure}

Because of the long, almost continuous observing campaign and the extreme variability of IRAS~13224, this is arguably the best dataset for studying the Fourier-domain properties of AGN variability. We have undertaken a major timing analysis \cite{Alston18}, which has produced several new exciting phenomena, including the detection of a quasi-periodic oscillation (QPO) feature in the power spectrum (Fig.~\ref{fig:timing_alston}). Unlike high-frequency QPOs detected in X-ray binaries and their analogues in AGN, the narrow feature in the IRAS~13224 power spectrum is strongest at low energies. This raises the possibility that it is instead due to ringing in the power spectrum, caused by X-ray reverberation, which has been predicted but not observed.

The other major result from this study is the discovery of a non-linear RMS-flux relation, for the first time in an accreting source (Fig.~\ref{fig:timing_alston}). The linear RMS-flux relation \cite{Uttley05_rms}, produced by multiplying together variations on different timescales, was thought to be ubiquitous in accreting sources. The reason for this deviation is not immediately apparent, but it potentially opens up a new regime for timing studies.

\section{Conclusions}
In this contribution, we have summarised key results from the 2016 observing campaign on IRAS~13224. This is arguably the best dataset for studying accretion physics in the strong gravity regime, and has lead to several major discoveries, both from spectroscopy and variability/timing studies, and is likely to give still more as we explore the data further. To date, the data suggest a scenario where a powerful wind is launched from the accretion disk at small radii, absorbing the X-ray continuum, regulating the accretion flow, and shielding the outer disk from the X-rays. While this source is a particularly extreme NLS1, there is nothing qualitatively different about it, suggesting that many other sources will show the same phenomenology when studied to the same level. The rapid variability, high luminosity, and strong outflow are consistent with the picture of NLS1s as rapidly growing lower mass black holes, analogous to early universe quasars.

\providecommand{\href}[2]{#2}\begingroup\raggedright\endgroup



\end{document}